\RequirePackage{fix-cm}

\documentclass[%
twocolumn, %
nofootinbib,
]{revtex4-1}
\usepackage{cmap}

\usepackage{ucs}
\usepackage[utf8x]{inputenc}
\usepackage[T1,T2A]{fontenc}
\usepackage[german,russian,english]{babel}

\usepackage[sort&compress]{natbib}
\usepackage{amsmath}
\usepackage{amssymb}

\usepackage{mathtools}
\mathtoolsset{
showonlyrefs,
mathic = true
}

\allowdisplaybreaks

\usepackage{hyperref}
\hypersetup{backref,
 colorlinks=false}
\hypersetup{pdfborder=0 0 0}

\usepackage{breakurl}

\usepackage{microtype}
\UseMicrotypeSet[protrusion]{alltext}

\usepackage{fixltx2e}

\usepackage{nicefrac}

\usepackage{physymb}

\makeatletter
\def\ps@pprintTitle{%
     \let\@oddhead\@empty
     \let\@evenhead\@empty
     \let\@oddfoot\@empty
     \let\@evenfoot\@oddfoot}
\makeatother

\newcommand{\crd}[1]{\underline{\vphantom{j}{#1}}}

\begin{document}

\title{The simplest geometrization of Maxwell's equations}

\author{D. S. Kulyabov}
\email{yamadharma@gmail.com}
\author{A. V. Korolkova}
\email{avkorolkova@gmail.com} 
\author{L. A. Sevastyanov}
\email{leonid.sevast@gmail.com}

\affiliation{Telecommunication System Department\\
Peoples' Friendship University of Russia\\
Miklukho-Maklaya str., 6, Moscow, 117198, Russia}


\thanks{Sources:
  \url{https://bitbucket.org/yamadharma/articles-2013-geom-maxwell}}

\begin{abstract}







  For research in the field of transformation optics and for the
  calculation of optically inhomogeneous lenses the method of
  geometrization of the Maxwell equations seems to be perspective. The
  basic idea is to transform the coefficients of material equations,
  namely the dielectric permittivity and magnetic permeability in the
  effective geometry of space-time (besides the vacuum Maxwell
  equations). This allows us to solve the direct and inverse problems,
  that is, to find the permittivity and magnetic permeability for a
  given effective geometry (paths of rays), as well as finding an
  effective geometry on the dielectric permittivity and magnetic
  permeability. The most popular naive geometrization was proposed by
  J. Plebanski. Under certain limitations it is quite good for solving
  relevant problems. It should be noted that in his paper only the
  resulting formulas and exclusively for Cartesian coordinate systems
  are given. In our work we conducted a detailed derivation of
  formulas for the naive geometrization of Maxwell's equations, and
  these formulas are written for an arbitrary curvilinear coordinate
  system. This work is a step toward building a complete covariant
  geometrization of the macroscopic Maxwell’s equations.

\end{abstract}


\maketitle

\section{Introduction}
\label{sec:intro}

Differential geometry was an important language of physics
XX-th century. Basic elements of it where developed within the general 
relativity theory. There is a desire to use its power in other
areas of physics, in particular in the optics.

The first attempts to apply the methods of differential geometry in
electrodynamics should be attributed to publications
I.~E.~Tamm~\cite{tamm:1924:jrpc::en, tamm:1925:jrpc::en,
   tamm:1925:mathann}. In 1960 E.~Plebansky proposed method of
geometrization of the constitutive equations of the electromagnetic field, which became
classic~\cite{Plebanski1960, Felice1971, Leonhardt2008,
   leonhardt:2009:light}. All subsequent works, either used it
or tried to correct a little, without changing
ideology~\cite{Thompson2011}. Unfortunately Plebansky~\cite{Plebanski1960} gives no
deriving formulas. Ideology is not expressed 
explicitly too.

For applying and deepening geometrization of material
equations the authors have restored the ideology and specific
Plebanski's calculations.

In paragraph~\ref{sec:notation} we prosecuted provides basic notation and conventions
used in the article. In paragraph~\ref{sec:maxwell_curv} are the main
relations for the Maxwell's equations in curvilinear coordinates (for
more detailed discussion the reader can be refer to other 
authors articles~\cite{kulyabov:2011:vestnik:curve-maxwell::en,
  kulyabov:2012:vestnik:2012-1}). In paragraph~\ref{sec:formal_geometr} are presented actual
calculations on Plebanski geometrization.

\section{Notations and conventions}
\label{sec:notation}

  \begin{enumerate}

  \item We will use the notation of abstract
    indices~\cite{penrose-rindler-1987::en}. In this notation tensor
    as a 
    complete object is denoted merely by an index (eg, $x^{i}$). Its
    components are
    designated by underlined indices (e.g., $x^{\crd{i}}$).

  \item We will adhere to the following agreements . Greek indices
    ($\alpha$, $\beta$) will refer to the four-dimensional space , in
    component form it looks like: $\crd{\alpha} = \overline{0,3}$.  Latin
    indices from the middle of the alphabet ($i$, $j$, $k$) will refer
    to the three-dimensional space , in the component form it looks like:
    $\crd {i} = \overline{1,3} $.

  \item The comma in the index denotes partial derivative with respect to
    corresponding coordinate ($f_{, i} := \partial_{i}f$); semicolon
    denotes covariant derivative ($f_{;i} := \nabla_{i} f$).

  \item To write the equations of electrodynamics in the article is used
    CGS symmetrical system~\cite{sivukhin:1979:ufn::en}.

  \item Antisymmetrization is denoted by straight brackets.

  \end{enumerate}

\section{Maxwell's equations in curvilinear coordinates}
\label{sec:maxwell_curv}

Here are the basics of Maxwell's equations in curvilinear
coordinates. A more detailed description is given in
articles~\cite{kulyabov:2012:vestnik:2012-1, kulyabov:2013:springer:cadabra,
  kulyabov:2013:conf:maxwell,
  kulyabov:2011:vestnik:curve-maxwell::en}.

Let's write the Maxwell equations with the help of electromagnetic
field tensors $F_{\alpha\beta}$ and
$G_{{\alpha}{\beta}}$~\cite{minkowski:1908,stratton:1948::en}:
\begin{gather}
\nabla_{{\alpha}} F_{{\beta}{\gamma}}+ \nabla_{{\beta}}
F_{{\gamma}{\alpha}}+\nabla_{{\gamma}} F_{{\alpha}{\beta}} = 
F_{[\alpha \beta ; \gamma]} = 0,
\label{eq:m:tensor:2}
\\
\nabla_{{\alpha}} G^{{\alpha}{\beta}}=\frac{4 \pi}{c}j^{{\beta}},
\label{eq:m:tensor}
\end{gather}
where the tensors $F_{\alpha \beta}$ and $G^{\alpha \beta}$ have the following
components
\begin{gather}
F_{\crd{\alpha}\crd{\beta}}=
\begin{pmatrix} 
0 & {E}_1 & {E}_2 & {E}_3 \\ 
-{E}_1 & 0 & -{B}^3 & {B}^2 \\
-{E}_2 & {B}^3 & 0 & -{B}^1 \\ 
-{E}_3 & -{B}^2 & {B}^1 & 0 
\end{pmatrix},
\label{eq:f_ab}
\\
F^{\crd{\alpha}\crd{\beta}}=
\begin{pmatrix} 
0 & -{E}^1 & -{E}^2 & -{E}^3 \\ 
{E}^1 & 0 & -{B}_3 & {B}_2 \\
{E}^2 & {B}_3 & 0 & -{B}_1 \\ 
{E}^3 & -{B}_2 & {B}_1 & 0 
\end{pmatrix},
\label{eq:f^ab}
\\
G^{\crd{\alpha}\crd{\beta}}=
\begin{pmatrix} 
0 & -{D}^1 & -{D}^2 & -{D}^3 \\ 
{D}^1 & 0 & -{H}_3 & {H}_2 \\
{D}^2 & {H}_3 & 0 & -{H}_1 \\ 
{D}^3 & -{H}_2 & {H}_1 & 0 
\end{pmatrix},
\label{eq:g^ab}
\\
G_{\crd{\alpha}\crd{\beta}}=
\begin{pmatrix} 
0 & {D}_1 & {D}_2 & {D}_3 \\ 
-{D}_1 & 0 & -{H}^3 & {H}^2 \\
-{D}_2 & {H}^3 & 0 & -{H}^1 \\ 
-{D}_3 & -{H}^2 & {H}^1 & 0 
\end{pmatrix}.
\label{eq:g_ab}
\end{gather}
  Where $E_{\crd{i}}$, $H_{\crd{i}}$ are components of electric and magnetic
  fields intensity vectors; $D^{\crd{i}}$, $B^{\crd{i}}$ are
  components of vectors of electric and magnetic induction.

It is also useful to introduce the tensor $\prescript{*}{}{F}^{\alpha\beta}$ dual conjugated to
${F}_{\alpha\beta}$
\begin{equation}
  \label{eq:tilde_f}
  \prescript{*}{}{F}^{\alpha\beta} = \frac{1}{2} e^{\alpha\beta\gamma\delta} F_{\gamma\delta},
\end{equation}
where $e^{\alpha\beta\gamma\delta}$ is the alternating tensor,
$\varepsilon^{\alpha\beta\gamma\delta}$ is the Levi-Civita symbol:
\begin{equation}
  \label{eq:e4}
  e_{\crd{\alpha}\crd{\beta}\crd{\gamma}\crd{\delta}} = \sqrt{-g}\varepsilon_{\crd{\alpha}\crd{\beta}\crd{\gamma}\crd{\delta}},\quad
  e^{\crd{\alpha}\crd{\beta}\crd{\gamma}\crd{\delta}} = -\frac{1}{\sqrt{-g}}\varepsilon^{\crd{\alpha}\crd{\beta}\crd{\gamma}\crd{\delta}}.
\end{equation}

Similarly, we can write
\begin{equation}
  \label{eq:tilde_other}
    \prescript{*}{}{G}_{\alpha\beta} = \frac{1}{2}
    e_{\alpha\beta\gamma\delta} G^{\gamma\delta}, 
\end{equation}
The components of these dual tensors are given (by the expressions) as follows
\begin{gather}
\prescript{*}{}{F}^{\crd{\alpha}\crd{\beta}}=
\frac{1}{\sqrt{-g}}
\begin{pmatrix} 
0 & -{B}^1 & -{B}^2 & -{B}^3 \\ 
{B}^1 & 0 & {E}_3 & -{E}_2 \\
{B}^2 & -{E}_3 & 0 & {E}_1 \\ 
{B}^3 & {E}_2 & -{E}_1 & 0 
\end{pmatrix},
\label{eq:*f^ab}
\\
\prescript{*}{}{G}_{\crd{\alpha}\crd{\beta}}=
\sqrt{-g}
\begin{pmatrix} 
0 & {H}_1 & {H}_2 & {H}_3 \\ 
-{H}_1 & 0 & {D}^3 & -{D}^2 \\
-{H}_2 & -{D}^3 & 0 & {D}^1 \\ 
-{H}_3 & {D}^2 & -{D}^1 & 0 
\end{pmatrix}.
\label{eq:*g_ab}
\end{gather}

With dual tensor~\eqref{eq:tilde_f}
the equation \eqref{eq:m:tensor:2} may be rewritten in a simpler form:
\begin{equation}
  \label{eq:m:tensor:2:dual}
  \nabla_{\alpha} \prescript{*}{}{F}^{\alpha\beta} = 0.
\end{equation}

Next, we will explain why, for the purposes of implementation of the
Plebanski's program, we prefer to write equations in the
form~\eqref{eq:m:tensor:2} rather the in simpler
form~\eqref{eq:m:tensor:2:dual}.

\subsection{Maxwell's equations in a medium}

In the presence of the medium a group of equations
Maxwell containing bound charges is changed, namely
equation~\eqref{eq:m:tensor}.
Amongst the Maxwell's equations~\eqref{eq:m:tensor:2}
and~\eqref{eq:m:tensor} must be added the constitutive relations between
tensors $G^{\alpha \beta}$ and $F^{\alpha \beta}$. When introduced
additional assumptions about the linearity of the environment and immobility
substances they may be written in three-dimensional form as follows:
\begin{equation}
  \label{eq:constraint}
  D^{i} = \varepsilon^{i j} E_{j}, \qquad B^{i} = \mu^{i j} H_{j},
\end{equation}
where $\varepsilon^{ij}$ and $\mu^{ij}$ are the permittivity and
permeability tensors. In four-dimensional form relation~\eqref{eq:constraint}
takes the following form:
\begin{equation}
  \label{eq:constraint:4}
  G^{\alpha \beta} = \lambda^{\alpha \beta}_{\gamma \delta} F^{\gamma
    \delta}, 
  \quad 
  \lambda^{\alpha \beta}_{\gamma \delta} = \lambda^{[\alpha \beta]}_{[\gamma \delta]},
\end{equation}
here tensor $\lambda^{\alpha \beta}_{\gamma \delta}$ is containing
information on the permittivity and permeability, as well as
electro-magnetic coupling~\cite{tamm:1924:jrpc::en}.

Given the structure tensor $F^{\alpha \beta}$~\eqref{eq:f^ab} and
$G^{\alpha \beta}$~\eqref{eq:g^ab}, and the
constitutive relations~\eqref{eq:constraint}, we write
\begin{equation}
  \label{eq:f-g_lambda}
  \begin{gathered}
    F^{\crd{0}\crd{i}} = - E^{\crd{i}}, \quad
    G^{\crd{0}\crd{i}} = - D^{\crd{i}},
    \\
    G^{\crd{i}\crd{j}} = - \varepsilon^{\crd{i} \crd{j} \crd{k}}
    H_{\crd{k}}, \quad
    B_{\crd{i}} = - \varepsilon_{\crd{i} \crd{j} \crd{k}} F^{\crd{j}
      \crd{k}}.
  \end{gathered}
\end{equation}
Or in another form:
\begin{equation}
  \label{eq:f-g_lambda_2}
  G^{\crd{0}\crd{i}} = \varepsilon^{\crd{i}}_{\crd{j}} F^{0 \crd{j}},
  \quad
  G^{\crd{i}\crd{j}} = \varepsilon^{\crd{i} \crd{j} \crd{k}}
  \varepsilon_{\crd{l} \crd{m} \crd{n}}
  (\mu^{-1})^{\crd{l}}_{\crd{k}} F^{\crd{m} \crd{n}}/2.
\end{equation}

From the~\eqref{eq:f-g_lambda_2} we write the structure of tensor
$\lambda^{\alpha \beta}_{\gamma \delta}$:
\begin{equation}
  \label{eq:lambda_struct}
  \lambda^{0 \crd{i}}_{0 \crd{j}} = \varepsilon^{\crd{i}}_{\crd{j}} /2, \quad
  \lambda^{0 \crd{i}}_{\crd{k} \crd{l}} = \lambda_{0 \crd{i}}^{\crd{k}
    \crd{l}} = 0, \quad
  \lambda^{\crd{i} \crd{j}}_{\crd{m} \crd{n}} = 
  \varepsilon^{\crd{i} \crd{j} \crd{k}}
  \varepsilon_{\crd{l} \crd{m} \crd{n}}
  (\mu^{-1})^{\crd{l}}_{\crd{k}}/2.
\end{equation}

Assuming that the vacuum permittivity and permeability are
view:
\begin{equation}
  \label{eq:perv:vac}
  \varepsilon^{i j} := \delta^{i j}, \qquad \mu^{i j} := \delta^{i j},
\end{equation}
we find that the vacuum constitutive relations~\eqref{eq:constraint} and~\eqref{eq:constraint:4} take
view
\begin{equation}
  \label{eq:constraint:vac}
  D^{i} = E^{i}, \qquad B^{i} = H^{i}, 
  \qquad G^{\alpha \beta} = F^{\alpha \beta}.
\end{equation}

\subsubsection{Permeability tensor in an isotropic medium}
\label{sec:perm_isotropic}

In the case of an isotropic medium, the
expression~\eqref{eq:constraint} takes the form:
\begin{equation}
  \label{eq:constraint:isotropic}
  D^{i} = \varepsilon E^{j}, \qquad B^{i} = \mu H^{j},
\end{equation}
where $\varepsilon^{ij}$ and $\mu^{ij}$ are the permittivity and
permeability scalars.

In this case, the permeability tensor in a stationary frame of
reference can be represented as follows~\cite{tamm:1925:mathann,
  tamm:1924:jrpc::en}:
\begin{equation}
  \label{eq:perm_tensor_2val}
  \begin{gathered}
    \lambda_{\alpha \beta \gamma \delta} = \lambda_{\alpha \gamma}
    \lambda_{\beta \delta},
    \\
    \lambda_{\crd{\alpha} \crd{\beta}} = 
    \mathrm{diag}\left(
      \frac{1}{\varepsilon \sqrt{\mu}}, -\sqrt{\mu}, -\sqrt{\mu},
      -\sqrt{\mu}
      \right), 
      \\
    \lambda^{\crd{\alpha} \crd{\beta}} =
    \mathrm{diag}\left(
      \varepsilon \sqrt{\mu}, -\frac{1}{\sqrt{\mu}}, -\frac{1}{\sqrt{\mu}},
      -\frac{1}{\sqrt{\mu}}
      \right).
  \end{gathered}
\end{equation}

\subsubsection{Constitutive relations in moving media}
\label{sec:moving_media}

Minkowski derived equations for connection
isotropic moving media~\cite{minkowski:1908, sommerfeld1964lectures}
(Minkowski's equations for moving media).
Let $u^{\alpha}$ is a environments four-speed. Assuming that the
permittivity and permeability $\varepsilon$ and $\mu$ are
scalars, we can write
\begin{equation}
  \label{eq:moving_isotropic_4d}
  G^{\alpha \beta} u_{\beta} = \varepsilon F^{\alpha \beta} u_{\beta}, 
  \quad
  \prescript{*}{}{F}^{\alpha \beta} u_{\beta} = \mu 
  \prescript{*}{}{G}^{\alpha \beta} u_{\beta}.
\end{equation}

In three-dimensional form of the
equation~\eqref{eq:moving_isotropic_4d} take the following form:
\begin{equation}
  \label{eq:moving_isotropic_3d}
  \begin{gathered}
    \begin{multlined}
      D^{i} = \varepsilon \left( E^{i} + 
        \left[\frac{u_{j}}{c}, B_{k} \right]^{i}
      \right) -
      \left[\frac{u_{j}}{c}, H_{k} \right]^{i} 
      = {} \\ {} =
      \varepsilon E^{i} + (\varepsilon \mu - 1) 
      \left[\frac{u_{j}}{c}, H_{k} \right]^{i}, 
    \end{multlined}
    \\
    \begin{multlined}
      B^{i} = \mu \left( H^{i} - 
        \left[\frac{u_{j}}{c}, D_{k} \right]^{i}
      \right) +
      \left[\frac{u_{j}}{c}, E_{k} \right]^{i} 
      = {} \\ {} = 
      \mu H^{i} - (\varepsilon \mu - 1) 
      \left[ \frac{u_{j}}{c}, E_{k} \right]^{i}.
    \end{multlined}
  \end{gathered}
\end{equation}

Tamm expanded equation~\eqref{eq:moving_isotropic_3d} for the
anisotropic case~\cite{tamm:1924:jrpc::en,
  tamm:1925:mathann}, namely, assuming that the permittivity and
permeability are of the form
\begin{equation}
  \label{eq:perm_anisotropic}
  \varepsilon^{\crd{i}}_{\crd{j}} = 
  \mathrm{diag}(\varepsilon^{1}_{1}, \varepsilon^{2}_{2},
  \varepsilon^{3}_{3}),
  \quad
  \mu^{\crd{i}}_{\crd{j}} = 
  \mathrm{diag}(\mu^{1}_{1}, \mu^{2}_{2},
  \mu^{3}_{3}),
\end{equation}
and velocity vector $u^{i}$ of frame of reference is parallel to one
of the principal axes of anisotropy. Then the Minkowski's equations
for moving media acquire the following form:
\begin{equation}
  \label{eq:moving_anisotropic_3d}
  \begin{gathered}
    D^{i} = \varepsilon^{i}_{l} \left( E^{l} + 
      \left[\frac{u_{j}}{c}, B_{k} \right]^{l}
    \right) - 
    \left[\frac{u_{j}}{c}, H_{k} \right]^{i} 
    , 
    \\
    B^{i} = \mu^{i}_{l} \left( H^{l} - 
      \left[\frac{u_{j}}{c}, D_{k} \right]^{l}
    \right) + 
    \left[\frac{u_{j}}{c}, E_{k} \right]^{i} 
    .
  \end{gathered}
\end{equation}

\section{Formal geometrization of material Maxwell's equations}
\label{sec:formal_geometr}

Plebanski has offered the elementary geometrization of Maxwell's
equations~\cite{Plebanski1960, Felice1971}. However, in the original
article the final formula are immediately given, and the principles and
methods for their preparation remain obscure. The authors tried to
explicitly describe technique that we believe Plebanski used and to perform
calculations in detail.

The basic ideas of Plebanski's geometrization are as follows:
\begin{enumerate}
\item We write the Maxwell's equations in a medium in Minkowski space.
\item We write the vacuum Maxwell's equations in the effective Riemann space.
\item We equate the corresponding terms of both equations.
\end{enumerate}

As a result, we obtain an expression of the permittivity and
permeability in terms of geometric objects.

Before attempting the program of Plebanski let us recall some auxiliary relations.

\subsection{Auxiliary relations}

\subsubsection{Differential Bianchi identity}

Note that the equation~\eqref{eq:m:tensor:2} can be written as

\begin{equation}
\label{eq:m:tensor:2:partial}
\partial_{{\alpha}} F_{{\beta}{\gamma}}+ \partial_{{\beta}}
F_{{\gamma}{\alpha}}+\partial_{{\gamma}} F_{{\alpha}{\beta}} = 
F_{[\alpha \beta , \gamma]} = 0.
\end{equation}

Really:
\begin{multline}
\label{eq:m:tensor:2:detail}
\nabla_{{\alpha}} F_{{\beta}{\gamma}}+ \nabla_{{\beta}}
F_{{\gamma}{\alpha}}+\nabla_{{\gamma}} F_{{\alpha}{\beta}} = {} \\ {} =
\partial_{{\alpha}} F_{{\beta}{\gamma}} -
\Gamma_{\alpha\beta}^{\delta} F_{\delta \gamma} - 
\Gamma_{\alpha \gamma}^{\delta} F_{\beta \delta} +
\partial_{{\beta}} F_{{\gamma}{\alpha}} -
\Gamma_{\beta \gamma}^{\delta} F_{\delta \alpha} -
\Gamma_{\beta \alpha}^{\delta} F_{\gamma \delta}
+ {} \\ {} +
\partial_{{\gamma}} F_{{\alpha}{\beta}} -
\Gamma_{\gamma \alpha}^{\delta} F_{\delta \beta} -
\Gamma_{\gamma \beta}^{\delta} F_{\alpha \delta}.
\end{multline}
Taking into account the anti-symmetry of the tensor $F_{\alpha \beta}$
and the
symmetry in the lower indices of Christoffel symbols $\Gamma_{\alpha
  \beta}^{\delta}$, we obtain~\eqref{eq:m:tensor:2:partial}.

The resulting equation can be written form-invariant by in any
coordinate system. Therefore, we will use the
equation~\eqref{eq:m:tensor:2:partial} instead of the more commonly
used equation~\eqref{eq:m:tensor:2:dual}.

\subsubsection{The metric tensor relations}

In addition, we need simple relations for the metric
tensor.
\begin{equation}
  \label{eq:g_g^:0}
  g_{\alpha \delta}g^{\delta \beta} = \delta_{\alpha}^{\beta}
\end{equation}
The relation~\eqref{eq:g_g^:0} leads to the following special relations:
\begin{gather}
  g_{0 \delta}g^{\delta i} = g_{0 0}g^{0 i} + g_{0 k}g^{k i}  = \delta_{0}^{i} = 0,
  \label{eq:g_g^:2}
  \\
  g_{i \delta}g^{\delta j} = g_{i 0}g^{0 j} + g_{i k}g^{k j}
  = \delta_{i}^{j}.
  \label{eq:g_g^:3}
\end{gather}
Let equation~\eqref{eq:g_g^:2} be rewritten in
form
\begin{gather}
  g^{0 i} = - \frac{1}{g_{0 0}} g_{0 k}g^{k i}.
  \label{eq:g_g^:2a}
\end{gather}
Substituting equation~\eqref{eq:g_g^:2a} in
equation~\eqref{eq:g_g^:3}, we obtain:
\begin{gather}
  \left(
    g_{i k} - \frac{1}{g_{0 0}} g_{0 i} g_{0 k}
  \right) g^{k j}
  = \delta_{i}^{j}.
  \label{eq:g_g^:2b}
\end{gather}

This relation will be used later to simplify the final equations.

\subsection{Geometrization in Cartesian coordinates}

Let us write the Maxwell's equations in a medium in Cartesian
coordinates with the metric tensor $\eta_{\crd{\alpha} \crd{\beta}} =
\mathrm{diag}(1,-1,-1,-1)$:
\begin{equation}
    \label{eq:maxwell:media:decart}
    \begin{gathered}
      \partial_{{\alpha}} F_{{\beta}{\gamma}} + 
      \partial_{{\beta}} F_{{\gamma}{\alpha}} + 
      \partial_{{\gamma}} F_{{\alpha}{\beta}} = 0,
      \\
      \partial_{{\alpha}} G^{{\alpha}{\beta}} = 
      \frac{4 \pi}{c} j^{{\beta}}.
    \end{gathered}
  \end{equation}

Now we write the vacuum Maxwell's equations in effective Riemann space
with the metric tensor $g_{\alpha \beta}$ (their related values mark
on by the tilde):
\begin{equation}
    \label{eq:maxwell:vacuum:riemann}
    \begin{gathered}
      \partial_{{\alpha}} \Tilde{F}_{{\beta}{\gamma}} + 
      \partial_{{\beta}} \Tilde{F}_{{\gamma}{\alpha}} + 
      \partial_{{\gamma}} \Tilde{F}_{{\alpha}{\beta}} = 0,
      \\
      \frac{1}{\sqrt{-g}}
      \partial_{{\alpha}} 
      \left( \sqrt{-g} 
        \Tilde{G}^{{\alpha}{\beta}} 
      \right)
        = 
      \frac{4 \pi}{c} \tilde{j}^{{\beta}}.
    \end{gathered}
  \end{equation}

In a vacuum, the following relation is true
(see~\eqref{eq:constraint:vac}):
\begin{equation}
  \label{eq:f-g:relation}
  \Tilde{F}_{\alpha \beta} = \Tilde{G}_{\alpha \beta}.
\end{equation}

Raising the indices in~\eqref{eq:f-g:relation}, we obtain
\begin{equation}
  \label{eq:f^-g:relation}
  \Tilde{F}^{\alpha \beta} = g^{\alpha \gamma} g^{\beta \delta}
  \Tilde{G}_{\gamma \delta}.
\end{equation}

Comparing term by term~\eqref{eq:maxwell:media:decart}
and~\eqref{eq:maxwell:vacuum:riemann}, and taking into account
the~\eqref{eq:f^-g:relation} we obtain:
\begin{gather}
  \label{eq:decart-riemann:relation}
  F_{\alpha \beta} = \Tilde{F}_{\alpha \beta}, 
  \qquad
  j^{\alpha} = \sqrt{-g} \Tilde{j}^{\alpha},
  \\
  G^{\alpha \beta}
  = \sqrt{-g} g^{\alpha \gamma} g^{\beta \delta} F_{\gamma \delta}.
  \label{eq:decart-riemann:relation:g-f}
\end{gather}

Equations~\eqref{eq:decart-riemann:relation:g-f} are actually
4-dimensional geometrized constitutive relations~\eqref{eq:constraint:4},
we were seeking for. Following the Plebanski's program 
we must obtain the explicit form for the 3-dimensional constitutive
relations~\eqref{eq:constraint}.

\subsubsection{Electric displacement field}

Let us rewrite~\eqref{eq:decart-riemann:relation:g-f} in the form of:
\begin{equation}
  \label{eq:f-durch-g}
  F_{\alpha \beta} = 
  \frac{1}{\sqrt{-g}} g_{\alpha \gamma} g_{\beta \delta} 
  G^{\gamma \delta}
\end{equation}
and look for the value components of $F_{0 \crd{i}}$, taking into
account relations~\eqref{eq:f_ab} and~\eqref{eq:g^ab}:
\begin{multline}
  \label{eq:f_0i}
  F_{0 i} = E_{i} = \frac{1}{\sqrt{-g}} 
  g_{0 \gamma} g_{i \delta} G^{\gamma \delta} 
  = {} \\ {} = 
  \frac{1}{\sqrt{-g}} \left(
  g_{0 j} g_{i 0} G^{j 0} + g_{0 0} g_{i j} G^{0 j}
  \right) +
  \frac{1}{\sqrt{-g}} g_{0 j} g_{i k} G^{j k} 
  = {} \\ {} = 
  \frac{1}{\sqrt{-g}} 
  g_{0 0} \left(
    \frac{1}{g_{0 0}} g_{0 j} g_{i 0} - g_{i j} 
  \right) D^{j} -
  \frac{1}{\sqrt{-g}} g_{0 j} g_{i k} \varepsilon^{j k l} H_{l}.
\end{multline}
For induction components $D^{i}$ we apply the relation~\eqref{eq:g_g^:2b}
and obtain
\begin{equation}
  \label{eq:d^i}
  D^{i} = - \frac{\sqrt{-g}}{g_{0 0}} g^{i j} E_{j} + 
  \frac{1}{g_{0 0}} \varepsilon^{i j k} g_{j 0} H_{k}.
\end{equation}

From~\eqref{eq:d^i} we can formally deduce the expression for the
permittivity:
\begin{equation}
  \label{eq:e_ij} 
  \varepsilon^{i j} = - \frac{\sqrt{-g}}{g_{0 0}} g^{i j}.
\end{equation}
In this sense the second term in~\eqref{eq:d^i} needs further
clarification.

\subsubsection{Magnetic induction}

To obtain expressions for the magnetic induction we will
use tensors~\eqref{eq:*f^ab} and~\eqref{eq:*g_ab}.
Moving down the indices of $\prescript{*}{}{G}^{\alpha \beta}$
in terms of~\eqref{eq:decart-riemann:relation:g-f} and applying
relations \eqref{eq:tilde_f}, \eqref{eq:tilde_other} we obtain
\begin{equation}
  \label{eq:*g-durch-*f}
  \prescript{*}{}{G}_{\alpha \beta} = 
  \sqrt{-g} g_{\alpha \gamma} g_{\beta \delta} 
  \prescript{*}{}{F}^{\gamma \delta}.
\end{equation}

We will seek for the value components of $\prescript{*}{}{G}_{0
  \crd{i}}$:
\begin{multline}
  \label{eq:*g_0i}
  \prescript{*}{}{G}_{0 i} = \sqrt{-g} H_{i} = 
  \sqrt{-g} g_{0 \gamma} g_{i \delta} 
  \prescript{*}{}{F}^{\gamma \delta} 
  = {} \\ {} = 
  \sqrt{-g} \left(
  g_{0 j} g_{i 0} \prescript{*}{}{F}^{j 0} + 
  g_{0 0} g_{i j} \prescript{*}{}{F}^{0 j}
  \right) +
  \sqrt{-g} g_{0 j} g_{i k} 
  \prescript{*}{}{F}^{j k} 
  = {} \\ {} = 
  \sqrt{-g}
  g_{0 0} \left(
    \frac{1}{g_{0 0}} g_{0 j} g_{i 0} - g_{i j} 
  \right) \frac{1}{\sqrt{-g}} B^{j} 
  - {} \\ {} -
  \sqrt{-g} g_{0 j} g_{i k} \varepsilon^{j k l} 
  \frac{1}{\sqrt{-g}} E_{l}.
\end{multline}

Applying the relation~\eqref{eq:g_g^:2b} we obtain for $B^{i}$ the
following expression:
\begin{equation}
  \label{eq:b^i}
  B^{i} = - \frac{\sqrt{-g}}{g_{0 0}} g^{i j} H_{j} - 
  \frac{1}{g_{0 0}} \varepsilon^{i j k} g_{j 0} E_{k}.
\end{equation}

From~\eqref{eq:b^i} we can formally write the expression for
permeability:
\begin{equation}
  \label{eq:mu_ij} 
  \mu^{i j} = - \frac{\sqrt{-g}}{g_{0 0}} g^{i j}.
\end{equation}

Thus geometrized constitutive relations in Cartesian
coordinates are as follows:
\begin{equation}
  \label{eq:geom-maxwell:decart}
  \begin{gathered}
    D^{i} = \varepsilon^{i j} E_{j} + 
    \varepsilon^{i j k} w_{j} H_{k}, \\
    B^{i} = \mu^{i j} H_{j} - 
    \varepsilon^{i j k} w_{j} E_{k}, \\
    \varepsilon^{i j} = - \frac{\sqrt{-g}}{g_{0 0}} g^{i j}, \quad
    \mu^{i j} = - \frac{\sqrt{-g}}{g_{0 0}} g^{i j}, \quad 
    w_{i} = \frac{g_{i 0}}{g_{0 0}}.
  \end{gathered}
\end{equation}

These equations were given in the original
paper~\cite{Plebanski1960}. For now, we can assume that we have performed
our task.

\subsubsection{Electro-magnetic interaction term interpretation}

In the equation~\eqref{eq:geom-maxwell:decart} the term of
electro-magnetic interactions does
receive no interpretation in original Plebanski's article. However
Leonhard proposed to interpret it as a speed
of geometrized frame of reference~\cite {Leonhardt2008}. Indeed, on
the basis of~\eqref{eq:moving_isotropic_3d}
equation~\eqref{eq:geom-maxwell:decart} can be rewritten as:
\begin{equation}
  \label{eq:geom-maxwell:decart:moving}
  \begin{gathered}
    D^{i} = \varepsilon^{i j} E_{j} + 
    \left[ \frac{u_{j}}{c}, H_{k} \right]^{i}, \\
    B^{i} = \mu^{i j} H_{j} - 
    \left[ \frac{u_{j}}{c}, E_{k} \right]^{i}, \\
    \varepsilon^{i j} = - \frac{\sqrt{-g}}{g_{0 0}} g^{i j}, \quad
    \mu^{i j} = - \frac{\sqrt{-g}}{g_{0 0}} g^{i j}, \quad 
    u_{i} = \frac{g_{i 0}}{g_{0 0}} \frac{c \sqrt{g^{(3)}}}{n^2 - 1},
  \end{gathered}
\end{equation}
where $u^{i}$ is three-dimensional velocity of frame of reference,
$u^{i}$ is determinant of the spatial part of
metric tensor $g_{\alpha \beta}$,
$n = \sqrt{\varepsilon \mu}$ is refractive index.

\subsection{Geometrization in curvilinear coordinates}

We now extend the scope of the formulas obtained to the Maxwell's
equations in arbitrary curvilinear
coordinates. Suppose that this space is defined by the metric
tensor $\gamma_{\alpha \beta}$. Then the
system~\eqref{eq:maxwell:media:decart} takes the following form:
\begin{equation}
    \label{eq:maxwell:media:curve}
    \begin{gathered}
      \partial_{{\alpha}} F_{{\beta}{\gamma}} + 
      \partial_{{\beta}} F_{{\gamma}{\alpha}} + 
      \partial_{{\gamma}} F_{{\alpha}{\beta}} = 0,
      \\
      \frac{1}{\sqrt{-\gamma}} 
      \partial_{{\alpha}} \left(
      \sqrt{-\gamma} G^{{\alpha}{\beta}} 
      \right)
      = 
      \frac{4 \pi}{c} j^{{\beta}}.
    \end{gathered}
  \end{equation}

Further, repeating the steps for the effective Riemann
space~\eqref{eq:maxwell:vacuum:riemann},
\eqref{eq:f-g:relation}, \eqref{eq:f^-g:relation}, we obtain the
analogues to~\eqref{eq:decart-riemann:relation}
and~\eqref{eq:decart-riemann:relation:g-f} as follows
\begin{gather}
  \label{eq:curv-riemann:relation}
  F_{\alpha \beta} = \Tilde{F}_{\alpha \beta}, 
  \qquad 
  j^{\alpha} = \frac{\sqrt{-g}}{\sqrt{-\gamma}} \Tilde{j}^{\alpha},
  \\
  G^{\alpha \beta}
  = \frac{\sqrt{-g}}{\sqrt{-\gamma}} g^{\alpha \gamma} g^{\beta \delta} F_{\gamma \delta}.
  \label{eq:curv-riemann:relation:g-f}
\end{gather}

\subsubsection{Electric displacement field}

We write the expression~\eqref{eq:curv-riemann:relation:g-f} as:
\begin{equation}
  \label{eq:f-durch-g:curv}
  F_{\alpha \beta} = 
  \frac{\sqrt{-\gamma}}{\sqrt{-g}} g_{\alpha \gamma} g_{\beta \delta} 
  G^{\gamma \delta}.
\end{equation}

Arguing similarly to~\eqref{eq:f_0i}, we obtain for the components of the
electric displacement field $D^{i}$ the relation:
\begin{equation}
  \label{eq:d^i:curv}
  D^{i} = - \frac{\sqrt{-g}}{\sqrt{-\gamma}}\frac{1}{g_{0 0}} g^{i j} E_{j} + 
  \frac{1}{g_{0 0}} \varepsilon^{i j k} g_{j 0} H_{k}, 
\end{equation}
and the expression for the permittivity takes the form:
\begin{equation}
  \label{eq:e_ij:curv} 
  \varepsilon^{i j} = - \frac{\sqrt{-g}}{\sqrt{-\gamma}} \frac{1}{g_{0 0}} g^{i j}.
\end{equation}

\subsubsection{Magnetic induction}

Let's rewrite~\eqref{eq:*g-durch-*f}
taking into account~\eqref{eq:curv-riemann:relation:g-f}
\begin{equation}
  \label{eq:*g-durch-*f:curv}
  \prescript{*}{}{G}_{\alpha \beta} = 
  \frac{\sqrt{-g}}{\sqrt{\gamma}} g_{\alpha \gamma} g_{\beta \delta} 
  \prescript{*}{}{F}^{\gamma \delta}.
\end{equation}

By analogy with~\eqref{eq:*g_0i}, \eqref{eq:b^i} and~\eqref{eq:mu_ij}
we obtain relations for the $B^{i}$:
\begin{equation}
  \label{eq:b^i:curv}
  B^{i} = - \frac{\sqrt{-g}}{\sqrt{\gamma}}\frac{1}{g_{0 0}} g^{i j} H_{j} - 
  \frac{1}{g_{0 0}} \varepsilon^{i j k} g_{j 0} E_{k},
\end{equation}
and permeability takes the form
\begin{equation}
  \label{eq:mu_ij:curv} 
  \mu^{i j} = - \frac{\sqrt{-g}}{\sqrt{-\gamma}}\frac{1}{g_{0 0}} g^{i j}.
\end{equation}

Thus geometrized constitutive relations in curvilinear
coordinates with the metric tensor $\gamma_{\alpha \beta}$ are of
the following form:
\begin{equation}
  \label{eq:geom-maxwell:curv}
  \begin{gathered}
    D^{i} = \varepsilon^{i j} E_{j} + 
    \varepsilon^{i j k} w_{j} H_{k}, \\
    B^{i} = \mu^{i j} H_{j} - 
    \varepsilon^{i j k} w_{j} E_{k}, \\
    \varepsilon^{i j} = - \frac{\sqrt{-g}}{\sqrt{-\gamma}} \frac{1}{g_{0 0}} g^{i j}, \qquad
    \mu^{i j} = - \frac{\sqrt{-g}}{\sqrt{-\gamma}}\frac{1}{g_{0 0}} g^{i j}, \qquad 
    w_{i} = \frac{g_{i 0}}{g_{0 0}}.
  \end{gathered}
\end{equation}

\section{Conclusions}
\label{sec:concl}

The authors have restored  the program and calculations of Plebanski.
This approach to geometrization appears inconclusive. 
Namely, the very method of obtaining relations for geometrization
(eq.~\eqref{eq:f^-g:relation}) implicitly assumes an
isotropic medium~\eqref{eq:perm_tensor_2val}.
But this does
not prevent on to
use this method for the calculations in the transformational
optics.

\bibliographystyle{abbrvnat}

\bibliography{bib/geom-maxwell/geom-maxwell,bib/geom-maxwell/maxwell-curvecoord,bib/geom-maxwell/maxwell-curvecoord-other}


\end{document}


\graphicspath{{image/geom-maxwell/}}

\title{Простейшая геометризация уравнений Максвелла}

\author{Д. С. Кулябов}
\email{yamadharma@gmail.com}
\author{А. В. Королькова}
\email{avkorolkova@gmail.com} 
\author{Л. А. Севастьянов}
\email{leonid.sevast@gmail.com}

\affiliation{Кафедра систем телекоммуникаций\\
  Российский университет дружбы народов\\
  ул. Миклухо-Маклая, д.6, Москва, 117198, Россия}


\thanks{Исходные тексты:
  \url{https://bitbucket.org/yamadharma/articles-2013-geom-maxwell}}

\begin{abstract}
  
    






  Для проведения разработок в области трансформационной оптики и для
  расчёта линз перспективным представляется метод геометризации
  уравнений Максвелла. Основная идея заключается в переводе
  материальных уравнений Максвелла, а именно диэлектрической и
  магнитной проницаемости, в эффективную геометрию
  пространства-времени (и вакуумные уравнения Максвелла). Это позволит
  решать прямую и обратную задачи, то есть находить диэлектрическую и
  магнитную проницаемость по заданной эффективной геометрии (по
  траекториям лучей), а также находить эффективную геометрию по
  диэлектрической и магнитной проницаемости. Наиболее популярная
  наивная геометризация была предложена Плебанским. При определённых
  ограничениях она достаточно хорошо решает задачи в своей
  области. Следует отметить, что в оригинальной статье приводятся лишь
  результирующие формулы и исключительно для декартовых систем
  координат. В работе авторов проводится подробный вывод формул для
  наивной геометризации уравнений Максвелла, кроме того формулы
  выписываются для произвольной криволинейной системы
  координат. Данная работа рассматривается как этап для построения
  полной ковариантной геометризации макроскопических уравнений
  Максвелла.

\end{abstract}

  \keywords{уравнения Максвелла; материальные уравнения Максвелла;
    геометризация уравнений Максвелла; риманова геометрия;
    криволинейные координаты;
    геометризация Плебанского}

\maketitle

\section{Введение}
\label{sec:intro}

Аппарат дифференциальной геометрии являлся основным языком физики
XX-го века. Его базовые элементы развивались в рамках общей теории
относительности. Возникает желание применить этот развитый и к другим
областям физики, в частности к оптике.

Первые попытки применения методов дифференциальной геометрии в
электродинамике следует отнести к публикациям
И.~Е.~Тамма~\cite{tamm:1924:jrpc, tamm:1925:jrpc,
  tamm:1925:mathann}. В 1960~году Е.~Плебанский предложил метод
геометризации материальных уравнений электромагнитного
поля~\cite{Plebanski1960, Felice1971, Leonhardt2008,
  leonhardt:2009:light}, ставший
классическим. Все последующие работы либо использовали его,
либо пытались немного подправить, не меняя
идеологии~\cite{Thompson2011}. К сожалению, в статье Плебанского~\cite{Plebanski1960} нет
никакого вывода формул, а идеология вывода также не выражена
явно.

Для применения и углубления направления геометризации материальных
уравнений авторам потребовалось восстановить идеологию и 
вывод уравнений по методу, предложенному 
Плебанским. 

В~разделе~\ref{sec:notation} даются основные обозначения и соглашения,
применяемые в статье. В разделе~\ref{sec:maxwell_curv} вводятся основные
соотношения для уравнений Максвелла в криволинейных координатах (для
более подробного ознакомления можно обратиться к другим статьям
авторов~\cite{kulyabov:2011:vestnik:curve-maxwell,
  kulyabov:2012:vestnik:2012-1}). В разделе~\ref{sec:formal_geometr}
приводятся собственно
расчёты по геометризации Плебанского.

\section{Обозначения и соглашения}
\label{sec:notation}

  \begin{enumerate}

  \item Будем использовать нотацию абстрактных
    индексов~\cite{penrose-rindler-1987}. В данной нотации тензор как
    целостный объект обозначается просто индексом (например, $x^{i}$),
    компоненты обозначаются подчёркнутым индексом (например,
    $x^{\crd{i}}$).

  \item Будем придерживаться следующих соглашений.  Греческие индексы
    ($\alpha$, $\beta$) будут относиться к четырёхмерному
    пространству и в компонентном виде будут иметь следующие значения:
    $\crd{\alpha} = \overline{0,3}$. Латинские индексы из середины
    алфавита ($i$, $j$, $k$) будут относиться к трёхмерному
    пространству и в компонентном виде будут иметь следующие значения:
    $\crd{i} = \overline{1,3}$.

  \item Запятой в индексе обозначается частная производная по
    соответствующей координате ($f_{,i} := \partial_{i} f$);
    точкой с запятой --- ковариантная производная ($f_{;i} := \nabla_{i}
    f$).

  \item Для записи уравнений электродинамики в работе используется
    система СГС симметричная~\cite{sivukhin:1979:ufn}.

  \item Антисимметризация обозначается квадратными скобками.

  \end{enumerate}

\section{Уравнения Максвелла в криволинейных координатах}
\label{sec:maxwell_curv}

\subsection{Общие соотношения}

Приведём основные сведения об уравнениях Максвелла в криволинейных
координатах. Более подробное описание дано в
статьях~\cite{kulyabov:2012:vestnik:2012-1,
  kulyabov:2013:springer:cadabra, kulyabov:2013:conf:maxwell,
  kulyabov:2013:tver:maxwell, 
  kulyabov:2011:vestnik:curve-maxwell}. 

Запишем уравнение Максвелла через тензоры
электромагнитного поля 
$F_{\alpha\beta}$ и
$G_{{\alpha}{\beta}}$~\cite{minkowski:1908,stratton:1948, terletskiy-rybakov-1990}:
\begin{gather}
\nabla_{{\alpha}} F_{{\beta}{\gamma}}+ \nabla_{{\beta}}
F_{{\gamma}{\alpha}}+\nabla_{{\gamma}} F_{{\alpha}{\beta}} = 
F_{[\alpha \beta ; \gamma]} = 0,
\label{eq:m:tensor:2}
\\
\nabla_{{\alpha}} G^{{\alpha}{\beta}}=\frac{4 \pi}{c}j^{{\beta}},
\label{eq:m:tensor}
\end{gather}
где тензоры $F_{\alpha\beta}$ и $G^{\alpha\beta}$ имеют следующие
компоненты
\begin{gather}
F_{\crd{\alpha}\crd{\beta}}=
\begin{pmatrix} 
0 & {E}_1 & {E}_2 & {E}_3 \\ 
-{E}_1 & 0 & -{B}^3 & {B}^2 \\
-{E}_2 & {B}^3 & 0 & -{B}^1 \\ 
-{E}_3 & -{B}^2 & {B}^1 & 0 
\end{pmatrix},
\label{eq:f_ab}
\\
F^{\crd{\alpha}\crd{\beta}}=
\begin{pmatrix} 
0 & -{E}^1 & -{E}^2 & -{E}^3 \\ 
{E}^1 & 0 & -{B}_3 & {B}_2 \\
{E}^2 & {B}_3 & 0 & -{B}_1 \\ 
{E}^3 & -{B}_2 & {B}_1 & 0 
\end{pmatrix},
\label{eq:f^ab}
\\
G^{\crd{\alpha}\crd{\beta}}=
\begin{pmatrix} 
0 & -{D}^1 & -{D}^2 & -{D}^3 \\ 
{D}^1 & 0 & -{H}_3 & {H}_2 \\
{D}^2 & {H}_3 & 0 & -{H}_1 \\ 
{D}^3 & -{H}_2 & {H}_1 & 0 
\end{pmatrix},
\label{eq:g^ab}
\\
G_{\crd{\alpha}\crd{\beta}}=
\begin{pmatrix} 
0 & {D}_1 & {D}_2 & {D}_3 \\ 
-{D}_1 & 0 & -{H}^3 & {H}^2 \\
-{D}_2 & {H}^3 & 0 & -{H}^1 \\ 
-{D}_3 & -{H}^2 & {H}^1 & 0 
\end{pmatrix}.
\label{eq:g_ab}
\end{gather}
  Здесь $E_{\crd{i}}$, $H_{\crd{i}}$~--- компоненты векторов напряжённости
  электрического и магнитного полей соответственно; $D^{\crd{i}}$,
  $B^{\crd{i}}$~--- компоненты векторов электрической и магнитной
  индукции соответственно.

Также полезно ввести тензор $\prescript{*}{}{F}^{\alpha\beta}$, дуально сопряжённый
тензору ${F}_{\alpha\beta}$:
\begin{equation}
  \label{eq:tilde_f}
  \prescript{*}{}{F}^{\alpha\beta} = \frac{1}{2} e^{\alpha\beta\gamma\delta} F_{\gamma\delta},
\end{equation}
где $e^{\alpha\beta\gamma\delta}$~--- альтернирующий тензор,
выражающийся через символ Леви-Чивиты
$\varepsilon^{\alpha\beta\gamma\delta}$:
\begin{equation}
  \label{eq:e4}
  e_{\crd{\alpha}\crd{\beta}\crd{\gamma}\crd{\delta}} = \sqrt{-g}\varepsilon_{\crd{\alpha}\crd{\beta}\crd{\gamma}\crd{\delta}},\quad
  e^{\crd{\alpha}\crd{\beta}\crd{\gamma}\crd{\delta}} = -\frac{1}{\sqrt{-g}}\varepsilon^{\crd{\alpha}\crd{\beta}\crd{\gamma}\crd{\delta}}.
\end{equation}

Аналогично вводится тензор
\begin{equation}
  \label{eq:tilde_other}
    \prescript{*}{}{G}_{\alpha\beta} = \frac{1}{2}
    e_{\alpha\beta\gamma\delta} G^{\gamma\delta}, 
\end{equation}
Оба они записываются в компонентах в следующем виде:
\begin{gather}
\prescript{*}{}{F}^{\crd{\alpha}\crd{\beta}}=
\frac{1}{\sqrt{-g}}
\begin{pmatrix} 
0 & -{B}^1 & -{B}^2 & -{B}^3 \\ 
{B}^1 & 0 & {E}_3 & -{E}_2 \\
{B}^2 & -{E}_3 & 0 & {E}_1 \\ 
{B}^3 & {E}_2 & -{E}_1 & 0 
\end{pmatrix},
\label{eq:*f^ab}
\\
\prescript{*}{}{G}_{\crd{\alpha}\crd{\beta}}=
\sqrt{-g}
\begin{pmatrix} 
0 & {H}_1 & {H}_2 & {H}_3 \\ 
-{H}_1 & 0 & {D}^3 & -{D}^2 \\
-{H}_2 & -{D}^3 & 0 & {D}^1 \\ 
-{H}_3 & {D}^2 & -{D}^1 & 0 
\end{pmatrix}.
\label{eq:*g_ab}
\end{gather}

С помощью дуального тензора~\eqref{eq:tilde_f} уравнение \eqref{eq:m:tensor:2} можно
записать в более простом виде:
\begin{equation}
  \label{eq:m:tensor:2:dual}
  \nabla_{\alpha} \prescript{*}{}{F}^{\alpha\beta} = 0.
\end{equation}

Далее мы объясним, почему для целей реализации методики Плебанского
мы предпочитаем запись уравнений в виде~\eqref{eq:m:tensor:2}
более простому виду~\eqref{eq:m:tensor:2:dual}.

\subsection{Уравнения Максвелла в среде}

При наличии среды претерпевает изменения лишь группа уравнений
Максвелла, содержащая связанные заряды, а именно
уравнение~\eqref{eq:m:tensor}. 
Помимо самих уравнений Максвелла~\eqref{eq:m:tensor:2}
и~\eqref{eq:m:tensor} необходимо добавить уравнения связи между
тензорами $G^{\alpha \beta}$ и $F^{\alpha \beta}$. При введении
дополнительных предположений о линейности среды и неподвижности
вещества их можно записать в трёхмерном виде следующим образом:
\begin{equation}
  \label{eq:constraint}
  D^{i} = \varepsilon^{i j} E_{j}, \qquad B^{i} = \mu^{i j} H_{j},
\end{equation}
где $\varepsilon^{i j}$ и $\mu^{i j}$~--- тензоры диэлектрической и
магнитной проницаемостей. В четырёхмерном виде~\eqref{eq:constraint}
принимает следующий вид:
\begin{equation}
  \label{eq:constraint:4}
  G^{\alpha \beta} = \lambda^{\alpha \beta}_{\gamma \delta} F^{\gamma
    \delta}, 
  \quad 
  \lambda^{\alpha \beta}_{\gamma \delta} = \lambda^{[\alpha \beta]}_{[\gamma \delta]},
\end{equation}
здесь $\lambda^{\alpha \beta}_{\gamma \delta}$~--- тензор
приницаемостей, содержащий информацию как об диэлектрической и магнитной
проницаемостях, так и об электро-магнитной
связи~\cite{tamm:1924:jrpc, tamm:1925:mathann}.

Учитывая структуру тензоров $F^{\alpha \beta}$~\eqref{eq:f^ab} и
$G^{\alpha \beta}$~\eqref{eq:g^ab}, а
также уравнения связи~\eqref{eq:constraint}, запишем 
\begin{equation}
  \label{eq:f-g_lambda}
  \begin{gathered}
    F^{\crd{0}\crd{i}} = - E^{\crd{i}}, \quad
    G^{\crd{0}\crd{i}} = - D^{\crd{i}},
    \\
    G^{\crd{i}\crd{j}} = - \varepsilon^{\crd{i} \crd{j} \crd{k}}
    H_{\crd{k}}, \quad
    B_{\crd{i}} = - \varepsilon_{\crd{i} \crd{j} \crd{k}} F^{\crd{j}
      \crd{k}}.
  \end{gathered}
\end{equation}
Или в другом виде:
\begin{equation}
  \label{eq:f-g_lambda_2}
  G^{\crd{0}\crd{i}} = \varepsilon^{\crd{i}}_{\crd{j}} F^{0 \crd{j}},
  \quad
  G^{\crd{i}\crd{j}} = \varepsilon^{\crd{i} \crd{j} \crd{k}}
  \varepsilon_{\crd{l} \crd{m} \crd{n}}
  (\mu^{-1})^{\crd{l}}_{\crd{k}} F^{\crd{m} \crd{n}}/2.
\end{equation}

Из~\eqref{eq:f-g_lambda_2} выпишем структуру
тензора $\lambda^{\alpha \beta}_{\gamma \delta}$:
\begin{equation}
  \label{eq:lambda_struct}
  \lambda^{0 \crd{i}}_{0 \crd{j}} = \varepsilon^{\crd{i}}_{\crd{j}} /2, \quad
  \lambda^{0 \crd{i}}_{\crd{k} \crd{l}} = \lambda_{0 \crd{i}}^{\crd{k}
    \crd{l}} = 0, \quad
  \lambda^{\crd{i} \crd{j}}_{\crd{m} \crd{n}} = 
  \varepsilon^{\crd{i} \crd{j} \crd{k}}
  \varepsilon_{\crd{l} \crd{m} \crd{n}}
  (\mu^{-1})^{\crd{l}}_{\crd{k}}/2.
\end{equation}

Считая, что в вакууме диэлектрическая и магнитная проницаемость имеют
вид:
\begin{equation}
  \label{eq:perv:vac}
  \varepsilon^{i j} := \delta^{i j}, \qquad \mu^{i j} := \delta^{i j},
\end{equation}
получим, что в вакууме уравнения связи~\eqref{eq:constraint} и~\eqref{eq:constraint:4} принимают
вид
\begin{equation}
  \label{eq:constraint:vac}
  D^{i} = E^{i}, \qquad B^{i} = H^{i}, 
  \qquad G^{\alpha \beta} = F^{\alpha \beta}.
\end{equation}

\subsubsection{Тензор проницаемостей для изотропной среды}
\label{sec:perm_isotropic}

В~случае изотропной среды выражение~\eqref{eq:constraint} принимает вид:
\begin{equation}
  \label{eq:constraint:isotropic}
  D^{i} = \varepsilon E^{j}, \qquad B^{i} = \mu H^{j},
\end{equation}
где диэлектрическая и
магнитная проницаемости $\varepsilon$ и $\mu$~--- скалярные
величины.

В~этом случае тензор проницаемостей 
в покоящейся системе отсчёта
можно представить в следующем
виде~\cite{tamm:1925:mathann, tamm:1924:jrpc}:
\begin{equation}
  \label{eq:perm_tensor_2val}
  \begin{gathered}
    \lambda_{\alpha \beta \gamma \delta} = \lambda_{\alpha \gamma}
    \lambda_{\beta \delta},
    \\
    \lambda_{\crd{\alpha} \crd{\beta}} = 
    \mathrm{diag}\left(
      \frac{1}{\varepsilon \sqrt{\mu}}, -\sqrt{\mu}, -\sqrt{\mu},
      -\sqrt{\mu}
      \right), 
      \\
    \lambda^{\crd{\alpha} \crd{\beta}} =
    \mathrm{diag}\left(
      \varepsilon \sqrt{\mu}, -\frac{1}{\sqrt{\mu}}, -\frac{1}{\sqrt{\mu}},
      -\frac{1}{\sqrt{\mu}}
      \right).
  \end{gathered}
\end{equation}

\subsubsection{Уравнения связи для движущихся сред}
\label{sec:moving_media}

Минковским были выведены уравнения связи для 
изотропных движущихся
сред~\cite{minkowski:1908, sommerfeld1958} (уравнения Минковского для
движущихся сред). 
Пусть $u^{\alpha}$~--- 4-скорость среды.
Считая диэлектрическую и
магнитную проницаемости $\varepsilon$ и $\mu$ скалярами, можно записать
\begin{equation}
  \label{eq:moving_isotropic_4d}
  G^{\alpha \beta} u_{\beta} = \varepsilon F^{\alpha \beta} u_{\beta}, 
  \quad
  \prescript{*}{}{F}^{\alpha \beta} u_{\beta} = \mu 
  \prescript{*}{}{G}^{\alpha \beta} u_{\beta}.
\end{equation}

В~трёхмерном виде уравнения~\eqref{eq:moving_isotropic_4d} принимают
следующий вид:
\begin{equation}
  \label{eq:moving_isotropic_3d}
  \begin{gathered}
    \begin{multlined}
      D^{i} = \varepsilon \left( E^{i} + 
        \left[\frac{u_{j}}{c}, B_{k} \right]^{i}
      \right) -
      \left[\frac{u_{j}}{c}, H_{k} \right]^{i} 
      = {} \\ {} =
      \varepsilon E^{i} + (\varepsilon \mu - 1) 
      \left[\frac{u_{j}}{c}, H_{k} \right]^{i}, 
    \end{multlined}
    \\
    \begin{multlined}
      B^{i} = \mu \left( H^{i} - 
        \left[\frac{u_{j}}{c}, D_{k} \right]^{i}
      \right) +
      \left[\frac{u_{j}}{c}, E_{k} \right]^{i} 
      = {} \\ {} = 
      \mu H^{i} - (\varepsilon \mu - 1) 
      \left[ \frac{u_{j}}{c}, E_{k} \right]^{i}.
    \end{multlined}
  \end{gathered}
\end{equation}

Тамм расширил уравнения~\eqref{eq:moving_isotropic_3d} для
анизотропного случая~\cite{tamm:1924:jrpc,
  tamm:1925:mathann}, а именно, считая, что диэлектрическая и магнитная
проницаемости имеют вид
\begin{equation}
  \label{eq:perm_anisotropic}
  \varepsilon^{\crd{i}}_{\crd{j}} = 
  \mathrm{diag}(\varepsilon^{1}_{1}, \varepsilon^{2}_{2},
  \varepsilon^{3}_{3}),
  \quad
  \mu^{\crd{i}}_{\crd{j}} = 
  \mathrm{diag}(\mu^{1}_{1}, \mu^{2}_{2},
  \mu^{3}_{3}),
\end{equation}
и~вектор скорость $u^{i}$ системы отсчёта параллельно одной из главных осей
анизотропии. Тогда уравнения Минковского для движущихся сред
приобретут следующий вид:
\begin{equation}
  \label{eq:moving_anisotropic_3d}
  \begin{gathered}
    D^{i} = \varepsilon^{i}_{l} \left( E^{l} + 
      \left[\frac{u_{j}}{c}, B_{k} \right]^{l}
    \right) - 
    \left[\frac{u_{j}}{c}, H_{k} \right]^{i} 
    , 
    \\
    B^{i} = \mu^{i}_{l} \left( H^{l} - 
      \left[\frac{u_{j}}{c}, D_{k} \right]^{l}
    \right) + 
    \left[\frac{u_{j}}{c}, E_{k} \right]^{i} 
    .
  \end{gathered}
\end{equation}

\section{Формальная геометризация материальных уравнений Максвелла}
\label{sec:formal_geometr}

Плебанским была предложена простейшая геометризация уравнений
Максвелла~\cite{Plebanski1960,Felice1971}. Однако в оригинальной статье сразу даны
финальные формулы, а принципы и методы их получения остаются
непрояснёнными. Авторы постарались явно выписать методику, которую по
нашему мнению использовал Плебанский, а также подробно провести вычисления.

Основная идея геометризации по Плебанскому заключается в следующем:
\begin{enumerate}
\item Записать уравнения Максвелла в среде в пространстве Минковского.
\item Записать вакуумные уравнения Максвелла в эффективном римановом
  пространстве.
\item Приравнять соответствующие члены уравнений.
\end{enumerate}

В~результате мы получим выражение диэлектрической и магнитной
проницаемостей через геометрические объекты. 

Прежде, чем приступать к реализации программы Плебанского, напомним
некоторые вспомогательные соотношения.

\subsection{Вспомогательные соотношения}

\subsubsection{Дифференциальное тождество Бьянки}

Заметим, что уравнение~\eqref{eq:m:tensor:2} можно записать в виде

\begin{equation}
\label{eq:m:tensor:2:partial}
\partial_{{\alpha}} F_{{\beta}{\gamma}}+ \partial_{{\beta}}
F_{{\gamma}{\alpha}}+\partial_{{\gamma}} F_{{\alpha}{\beta}} = 
F_{[\alpha \beta , \gamma]} = 0.
\end{equation}

Действительно:
\begin{multline}
\label{eq:m:tensor:2:detail}
\nabla_{{\alpha}} F_{{\beta}{\gamma}}+ \nabla_{{\beta}}
F_{{\gamma}{\alpha}}+\nabla_{{\gamma}} F_{{\alpha}{\beta}} = {} \\ {} =
\partial_{{\alpha}} F_{{\beta}{\gamma}} -
\Gamma_{\alpha\beta}^{\delta} F_{\delta \gamma} - 
\Gamma_{\alpha \gamma}^{\delta} F_{\beta \delta} +
\partial_{{\beta}} F_{{\gamma}{\alpha}} -
\Gamma_{\beta \gamma}^{\delta} F_{\delta \alpha} -
\Gamma_{\beta \alpha}^{\delta} F_{\gamma \delta}
+ {} \\ {} +
\partial_{{\gamma}} F_{{\alpha}{\beta}} -
\Gamma_{\gamma \alpha}^{\delta} F_{\delta \beta} -
\Gamma_{\gamma \beta}^{\delta} F_{\alpha \delta}.
\end{multline}
Учитывая антисимметрию тензора $F_{\alpha \beta}$ и симметрию по
нижним индексам символа Кристоффеля $\Gamma_{\alpha \beta}^{\delta}$,
мы и получим~\eqref{eq:m:tensor:2:partial}.

Полученное уравнение записывается форминвариантно в произвольной
системе координат. Поэтому мы и будем использовать
уравнение~\eqref{eq:m:tensor:2:partial} вместо более общеупотребительного
уравнения~\eqref{eq:m:tensor:2:dual}.

\subsubsection{Соотношения для метрического тензора}

Нам понадобятся простые соотношения для метрического
тензора. Выражение
\begin{equation}
  \label{eq:g_g^:0}
  g_{\alpha \delta}g^{\delta \beta} = \delta_{\alpha}^{\beta}
\end{equation}
приводит к следующим частным соотношениям:
\begin{gather}
  g_{0 \delta}g^{\delta i} = g_{0 0}g^{0 i} + g_{0 k}g^{k i}  = \delta_{0}^{i} = 0,
  \label{eq:g_g^:2}
  \\
  g_{i \delta}g^{\delta j} = g_{i 0}g^{0 j} + g_{i k}g^{k j}
  = \delta_{i}^{j}.
  \label{eq:g_g^:3}
\end{gather}
Соотношение~\eqref{eq:g_g^:2} перепишем в
виде
\begin{gather}
  g^{0 i} = - \frac{1}{g_{0 0}} g_{0 k}g^{k i}.
  \label{eq:g_g^:2a}
\end{gather}
Подставляя~\eqref{eq:g_g^:2a} в~\eqref{eq:g_g^:3},
получаем:
\begin{gather}
  \left(
    g_{i k} - \frac{1}{g_{0 0}} g_{0 i} g_{0 k}
  \right) g^{k j}
  = \delta_{i}^{j}.
  \label{eq:g_g^:2b}
\end{gather}

Это соотношение будет использовано позднее для упрощения записи
итоговых уравнений.

\subsection{Геометризация в декартовых координатах}

Запишем уравнения Максвелла в среде в декартовых координатах с
метрическим тензором $\eta_{\crd{\alpha} \crd{\beta}} =
\mathrm{diag}(1,-1,-1,-1)$:
\begin{equation}
    \label{eq:maxwell:media:decart}
    \begin{gathered}
      \partial_{{\alpha}} F_{{\beta}{\gamma}} + 
      \partial_{{\beta}} F_{{\gamma}{\alpha}} + 
      \partial_{{\gamma}} F_{{\alpha}{\beta}} = 0,
      \\
      \partial_{{\alpha}} G^{{\alpha}{\beta}} = 
      \frac{4 \pi}{c} j^{{\beta}}.
    \end{gathered}
  \end{equation}

Теперь запишем вакуумные уравнения Максвелла в эффективном римановом
пространстве с метрическим тензором $g_{\alpha \beta}$ (относящиеся к
ним величины пометим тильдой):
\begin{equation}
    \label{eq:maxwell:vacuum:riemann}
    \begin{gathered}
      \partial_{{\alpha}} \Tilde{F}_{{\beta}{\gamma}} + 
      \partial_{{\beta}} \Tilde{F}_{{\gamma}{\alpha}} + 
      \partial_{{\gamma}} \Tilde{F}_{{\alpha}{\beta}} = 0,
      \\
      \frac{1}{\sqrt{-g}}
      \partial_{{\alpha}} 
      \left( \sqrt{-g} 
        \Tilde{G}^{{\alpha}{\beta}} 
      \right)
        = 
      \frac{4 \pi}{c} \tilde{j}^{{\beta}}.
    \end{gathered}
  \end{equation}

В вакууме будет верно следующее соотношение
(см.~\eqref{eq:constraint:vac}):
\begin{equation}
  \label{eq:f-g:relation}
  \Tilde{F}_{\alpha \beta} = \Tilde{G}_{\alpha \beta}.
\end{equation}

Подымая индексы в~\eqref{eq:f-g:relation}, получим
\begin{equation}
  \label{eq:f^-g:relation}
  \Tilde{F}^{\alpha \beta} = g^{\alpha \gamma} g^{\beta \delta}
  \Tilde{G}_{\gamma \delta}.
\end{equation}

Сравнивая почленно~\eqref{eq:maxwell:media:decart}
и~\eqref{eq:maxwell:vacuum:riemann}, с учётом~\eqref{eq:f^-g:relation}
получим:
\begin{gather}
  \label{eq:decart-riemann:relation}
  F_{\alpha \beta} = \Tilde{F}_{\alpha \beta}, 
  \qquad
  j^{\alpha} = \sqrt{-g} \Tilde{j}^{\alpha},
  \\
  G^{\alpha \beta}
  = \sqrt{-g} g^{\alpha \gamma} g^{\beta \delta} F_{\gamma \delta}.
  \label{eq:decart-riemann:relation:g-f}
\end{gather}

Уравнения~\eqref{eq:decart-riemann:relation:g-f} собственно и являются
искомыми 4-мерными геометризованными уравнениями
связи~\eqref{eq:constraint:4}. Следуя методике Плебанского мы должны
получить явный вид для трёхмерных уравнений связи~\eqref{eq:constraint}.

\subsubsection{Формула для вектора электрической индукции}

Перепишем выражение~\eqref{eq:decart-riemann:relation:g-f} в виде:
\begin{equation}
  \label{eq:f-durch-g}
  F_{\alpha \beta} = 
  \frac{1}{\sqrt{-g}} g_{\alpha \gamma} g_{\beta \delta} 
  G^{\gamma \delta}
\end{equation}
и будем искать значение компонент $F_{0 \crd{i}}$,
учитывая~\eqref{eq:f_ab} и \eqref{eq:g^ab}:
\begin{multline}
  \label{eq:f_0i}
  F_{0 i} = E_{i} = \frac{1}{\sqrt{-g}} 
  g_{0 \gamma} g_{i \delta} G^{\gamma \delta} 
  = {} \\ {} = 
  \frac{1}{\sqrt{-g}} \left(
  g_{0 j} g_{i 0} G^{j 0} + g_{0 0} g_{i j} G^{0 j}
  \right) +
  \frac{1}{\sqrt{-g}} g_{0 j} g_{i k} G^{j k} 
  = {} \\ {} = 
  \frac{1}{\sqrt{-g}} 
  g_{0 0} \left(
    \frac{1}{g_{0 0}} g_{0 j} g_{i 0} - g_{i j} 
  \right) D^{j} -
  \frac{1}{\sqrt{-g}} g_{0 j} g_{i k} \varepsilon^{j k l} H_{l}.
\end{multline}
Для компонент индукции $D^{i}$ применим соотношение~\eqref{eq:g_g^:2b}
и получим:
\begin{equation}
  \label{eq:d^i}
  D^{i} = - \frac{\sqrt{-g}}{g_{0 0}} g^{i j} E_{j} + 
  \frac{1}{g_{0 0}} \varepsilon^{i j k} g_{j 0} H_{k}.
\end{equation}

Из~\eqref{eq:d^i} можно формально выписать выражение для
диэлектрической проницаемости:
\begin{equation}
  \label{eq:e_ij} 
  \varepsilon^{i j} = - \frac{\sqrt{-g}}{g_{0 0}} g^{i j}.
\end{equation}
При этом смысл второго члена в~\eqref{eq:d^i} нуждается в дальнейшем уточнении.

\subsubsection{Формула для вектора магнитной индукции}

Для получения выражения для вектора магнитной индукции будем
использовать тензоры~\eqref{eq:*f^ab} и~\eqref{eq:*g_ab}. 
Опуская индексы у
$\prescript{*}{}{G}^{\alpha \beta}$
в выражении~\eqref{eq:decart-riemann:relation:g-f} и применяя
соотношения \eqref{eq:tilde_f} и \eqref{eq:tilde_other}, получаем:
\begin{equation}
  \label{eq:*g-durch-*f}
  \prescript{*}{}{G}_{\alpha \beta} = 
  \sqrt{-g} g_{\alpha \gamma} g_{\beta \delta} 
  \prescript{*}{}{F}^{\gamma \delta}.
\end{equation}

Будем искать значения компонент $\prescript{*}{}{G}_{0 \crd{i}}$:
\begin{multline}
  \label{eq:*g_0i}
  \prescript{*}{}{G}_{0 i} = \sqrt{-g} H_{i} = 
  \sqrt{-g} g_{0 \gamma} g_{i \delta} 
  \prescript{*}{}{F}^{\gamma \delta} 
  = {} \\ {} = 
  \sqrt{-g} \left(
  g_{0 j} g_{i 0} \prescript{*}{}{F}^{j 0} + 
  g_{0 0} g_{i j} \prescript{*}{}{F}^{0 j}
  \right) +
  \sqrt{-g} g_{0 j} g_{i k} 
  \prescript{*}{}{F}^{j k} 
  = {} \\ {} = 
  \sqrt{-g}
  g_{0 0} \left(
    \frac{1}{g_{0 0}} g_{0 j} g_{i 0} - g_{i j} 
  \right) \frac{1}{\sqrt{-g}} B^{j} 
  - {} \\ {} -
  \sqrt{-g} g_{0 j} g_{i k} \varepsilon^{j k l} 
  \frac{1}{\sqrt{-g}} E_{l}.
\end{multline}

Применяя соотношение~\eqref{eq:g_g^:2b}, получим для $B^{i}$ следующее
выражение:
\begin{equation}
  \label{eq:b^i}
  B^{i} = - \frac{\sqrt{-g}}{g_{0 0}} g^{i j} H_{j} - 
  \frac{1}{g_{0 0}} \varepsilon^{i j k} g_{j 0} E_{k}.
\end{equation}

Из~\eqref{eq:b^i} можно формально выписать выражение для
магнитной проницаемости:
\begin{equation}
  \label{eq:mu_ij} 
  \mu^{i j} = - \frac{\sqrt{-g}}{g_{0 0}} g^{i j}.
\end{equation}

Таким образом геометризованные уравнения связи в декартовых
координатах имеют следующий вид:
\begin{equation}
  \label{eq:geom-maxwell:decart}
  \begin{gathered}
    D^{i} = \varepsilon^{i j} E_{j} + 
    \varepsilon^{i j k} w_{j} H_{k}, \\
    B^{i} = \mu^{i j} H_{j} - 
    \varepsilon^{i j k} w_{j} E_{k}, \\
    \varepsilon^{i j} = - \frac{\sqrt{-g}}{g_{0 0}} g^{i j}, \quad
    \mu^{i j} = - \frac{\sqrt{-g}}{g_{0 0}} g^{i j}, \quad 
    w_{i} = \frac{g_{i 0}}{g_{0 0}}.
  \end{gathered}
\end{equation}

Эти уравнения и были приведены в исходной
статье~\cite{Plebanski1960}. Теперь мы можем считать
выполненной нашу задачу по восстановлению методики Плебанского.

\subsubsection{Интерпретация члена электро-магнитного взаимодействия}

В уравнениях~\eqref{eq:geom-maxwell:decart} член электро-магнитного
взаимодействия у Плебанского никакого истолкования не получает. Однако
Леонгард предложил интерпретировать его как скорость движения
геометризованной системы отсчёта~\cite{Leonhardt2008}. Действительно, 
на основании~\eqref{eq:moving_isotropic_3d}
уравнения~\eqref{eq:geom-maxwell:decart} можно переписать в виде:
\begin{equation}
  \label{eq:geom-maxwell:decart:moving}
  \begin{gathered}
    D^{i} = \varepsilon^{i j} E_{j} + 
    \left[ \frac{u_{j}}{c}, H_{k} \right]^{i}, \\
    B^{i} = \mu^{i j} H_{j} - 
    \left[ \frac{u_{j}}{c}, E_{k} \right]^{i}, \\
    \varepsilon^{i j} = - \frac{\sqrt{-g}}{g_{0 0}} g^{i j}, \quad
    \mu^{i j} = - \frac{\sqrt{-g}}{g_{0 0}} g^{i j}, \quad 
    u_{i} = \frac{g_{i 0}}{g_{0 0}} \frac{c \sqrt{g^{(3)}}}{n^2 - 1},
  \end{gathered}
\end{equation}
где $u^{i}$~--- трёхмерная скорость движения системы отсчёта, $g^{(3)}
= \mathrm{det} g_{ij}$~--- определитель пространственной части
метрического тензора $g_{\alpha \beta}$, $n = \sqrt{\varepsilon
  \mu}$~--- показатель преломления.

\subsection{Геометризация в криволинейных координатах}

Расширим теперь область применения полученных формул за счёт записи
уравнений Максвелла в произвольных криволинейных системах
координат. Пусть исходное пространство задаётся метрическим тензором
$\gamma_{\alpha \beta}$. Тогда система~\eqref{eq:maxwell:media:decart}
примет следующий вид:
\begin{equation}
    \label{eq:maxwell:media:curve}
    \begin{gathered}
      \partial_{{\alpha}} F_{{\beta}{\gamma}} + 
      \partial_{{\beta}} F_{{\gamma}{\alpha}} + 
      \partial_{{\gamma}} F_{{\alpha}{\beta}} = 0,
      \\
      \frac{1}{\sqrt{-\gamma}} 
      \partial_{{\alpha}} \left(
      \sqrt{-\gamma} G^{{\alpha}{\beta}} 
      \right)
      = 
      \frac{4 \pi}{c} j^{{\beta}}.
    \end{gathered}
  \end{equation}

Далее, повторяя шаги для эффективного риманового
пространства~\eqref{eq:maxwell:vacuum:riemann},
\eqref{eq:f-g:relation}, \eqref{eq:f^-g:relation}, получим
аналог~\eqref{eq:decart-riemann:relation}
и~\eqref{eq:decart-riemann:relation:g-f}:
\begin{gather}
  \label{eq:curv-riemann:relation}
  F_{\alpha \beta} = \Tilde{F}_{\alpha \beta}, 
  \qquad 
  j^{\alpha} = \frac{\sqrt{-g}}{\sqrt{-\gamma}} \Tilde{j}^{\alpha},
  \\
  G^{\alpha \beta}
  = \frac{\sqrt{-g}}{\sqrt{-\gamma}} g^{\alpha \gamma} g^{\beta \delta} F_{\gamma \delta}.
  \label{eq:curv-riemann:relation:g-f}
\end{gather}

\subsubsection{Формула для вектора электрической индукции}

Запишем выражение~\eqref{eq:curv-riemann:relation:g-f} в виде:
\begin{equation}
  \label{eq:f-durch-g:curv}
  F_{\alpha \beta} = 
  \frac{\sqrt{-\gamma}}{\sqrt{-g}} g_{\alpha \gamma} g_{\beta \delta} 
  G^{\gamma \delta}.
\end{equation}

Рассуждая аналогично~\eqref{eq:f_0i}, получим 
для компонент индукции $D^{i}$ соотношение:
\begin{equation}
  \label{eq:d^i:curv}
  D^{i} = - \frac{\sqrt{-g}}{\sqrt{-\gamma}}\frac{1}{g_{0 0}} g^{i j} E_{j} + 
  \frac{1}{g_{0 0}} \varepsilon^{i j k} g_{j 0} H_{k}, 
\end{equation}
а выражение для диэлектрической проницаемости примет вид:
\begin{equation}
  \label{eq:e_ij:curv} 
  \varepsilon^{i j} = - \frac{\sqrt{-g}}{\sqrt{-\gamma}} \frac{1}{g_{0 0}} g^{i j}.
\end{equation}

\subsubsection{Формула для вектора магнитной индукции}

Перепишем~\eqref{eq:*g-durch-*f} с
учётом~\eqref{eq:curv-riemann:relation:g-f}
\begin{equation}
  \label{eq:*g-durch-*f:curv}
  \prescript{*}{}{G}_{\alpha \beta} = 
  \frac{\sqrt{-g}}{\sqrt{\gamma}} g_{\alpha \gamma} g_{\beta \delta} 
  \prescript{*}{}{F}^{\gamma \delta}.
\end{equation}

По аналогии с~\eqref{eq:*g_0i}, \eqref{eq:b^i} и~\eqref{eq:mu_ij}
получим соотношение для $B^{i}$:
\begin{equation}
  \label{eq:b^i:curv}
  B^{i} = - \frac{\sqrt{-g}}{\sqrt{\gamma}}\frac{1}{g_{0 0}} g^{i j} H_{j} - 
  \frac{1}{g_{0 0}} \varepsilon^{i j k} g_{j 0} E_{k},
\end{equation}
и для магнитной проницаемости:
\begin{equation}
  \label{eq:mu_ij:curv} 
  \mu^{i j} = - \frac{\sqrt{-g}}{\sqrt{-\gamma}}\frac{1}{g_{0 0}} g^{i j}.
\end{equation}

Таким образом геометризованные уравнения связи в криволинейных
координатах с метрическим тензором $\gamma_{\alpha \beta}$ имеют
следующий вид:
\begin{equation}
  \label{eq:geom-maxwell:curv}
  \begin{gathered}
    D^{i} = \varepsilon^{i j} E_{j} + 
    \varepsilon^{i j k} w_{j} H_{k}, \\
    B^{i} = \mu^{i j} H_{j} - 
    \varepsilon^{i j k} w_{j} E_{k}, \\
    \varepsilon^{i j} = - \frac{\sqrt{-g}}{\sqrt{-\gamma}} \frac{1}{g_{0 0}} g^{i j}, \qquad
    \mu^{i j} = - \frac{\sqrt{-g}}{\sqrt{-\gamma}}\frac{1}{g_{0 0}} g^{i j}, \qquad 
    w_{i} = \frac{g_{i 0}}{g_{0 0}}.
  \end{gathered}
\end{equation}

\section{Выводы}
\label{sec:concl}

Авторами была восстановлена методика и расчёты Плебанского. Однако
данный подход к геометризации представляется неокончательным,
поскольку рассматривает специальный случай. А именно, сам метод
получения соотношений для геометризации
(уравнение~\eqref{eq:f^-g:relation}) неявно подразумевает изотропность
среды~\eqref{eq:perm_tensor_2val}.
Впрочем, это не мешает
применять данный метод при расчётах в области трансформационной
оптики.

\bibliographystyle{gost2008l}

\bibliography{bib/geom-maxwell/geom-maxwell,bib/geom-maxwell/maxwell-curvecoord,bib/geom-maxwell/maxwell-curvecoord-other}

